\documentclass[12pt,preprint]{aastex}

\shorttitle{First tests of multi-beacon wavefront sensing}
\shortauthors{Lloyd-Hart et al.}

\begin{document}

\title{First tests of wavefront sensing with a constellation of laser guide beacons}

\author{M. Lloyd-Hart, C. Baranec, N. M. Milton, T. Stalcup,
M. Snyder, N. Putnam, and J. R. P. Angel}
\affil{Steward Observatory, The University of Arizona,
Tucson, AZ 85721}

\email{mhart@as.arizona.edu}

\begin{abstract}

Adaptive optics to correct current telescopes over wide fields, or
future very large telescopes over even narrow fields, will require
real-time wavefront measurements made with a constellation of laser
beacons. Here we report the first such measurements, made at the 6.5 m
MMT with five Rayleigh beacons in a $2\arcmin$ pentagon. Each beacon
is made with a pulsed beam at 532 nm, of 4 W at the exit pupil of
the projector. The return is range-gated
from 20--29 km and recorded at 53 Hz by a 36-element Shack-Hartmann
sensor. Wavefronts derived from the beacons are compared with
simultaneous wavefronts obtained for individual natural stars within
or near the constellation. Observations were made in seeing averaging
$1\farcs0$
with 2/3 of the aberration measured to be from a ground layer of mean
height 380 m.  Under these conditions, subtraction of the simple
instantaneous average of the five beacon wavefronts from the stellar
wavefronts yielded a 40\% rms reduction in the measured modes of the
distortion over a $2\arcmin$ field. We discuss the use of multiple
Rayleigh beacons as an alternative to single sodium beacons on 8 m
telescopes, and the impact of the new work on the design of a
multi-sodium beacon system for the 25 m Giant Magellan Telescope.
\end{abstract}


\section{Introduction}

Ground based astronomical telescopes with adaptive correction are able
to exceed the resolution of the Hubble Space Telescope.  However, their
use is currently limited almost exclusively to a small fraction of the sky,
in small fields of view around the stars required to measure the wavefront
error.  These stars must be relatively bright because removal of atmospheric
blurring requires rapidly updated command signals to the wavefront corrector.
The radius of the compensated field around the star, the isoplanatic angle,
is limited because atmospheric
turbulence extends many kilometers above the ground, so the integrated
wavefront distortion depends sensitively on the direction of view.

For telescopes in the current 8--10 m class, the limitation on brightness,
though not on corrected field of view, can
be largely removed by use of a laser beam projected toward a faint astronomical
target to create an artificial beacon. Light from a sodium resonance beacon
generated at $\sim$ 90 km altitude follows the path taken by light from the
target closely enough that a good estimate of the aberrations in the latter
can be derived. The laser light, however, does not come from infinity, and as
the telescope diameter is increased, the mismatch in optical paths (the
cone effect) eventually becomes so severe that no useful recovery of stellar
wavefronts can be made.

Extremely large telescopes (ELTs) with diameters greater than 25 m
will therefore require a further advance to recover
diffraction-limited imaging with AO.    Multiple beacons placed so that
their light collectively samples the full volume of atmosphere traversed by
light from a target can in principle provide the solution.  The instantaneous
stellar wavefront would be computed by a tomographic algorithm applied to the
wavefronts measured from all the beacons \citep{Ragazzoni,TokTom,MLH03}.
This method, implemented with sodium beacon lasers, is planned for the
Giant Magellan Telescope (GMT) \citep{GMTAO}, and for the Thirty Meter
Telescope \citep{Dekany}, to recover the full resolution of their apertures. 

In the simplest application of tomography, the stellar wavefront along
a single line of sight is estimated from all the measured beacon wavefronts,
and the compensating phase is applied to a single deformable mirror (DM). The
technique, called Laser Tomography AO (LTAO), delivers a diffraction-limited
field of view limited by the normal isoplanatic angle.  

Once a telescope is equipped with multiple laser beacons and tomographic analysis,
a variety of altitude-conjugated AO techniques are enabled: diffraction-limited
resolution over a field many times the isoplanatic angle with multi-conjugate
and multi-object AO (MCAO, MOAO) \citep{Beckers,TokMCAO,Hammer,Dekany}, and
partially improved seeing over a yet larger field with ground-layer AO (GLAO).

Of these advanced techniques, only MCAO has yet been attempted, in experiments
at two solar telescopes \citep{Langlois,KIS}.  Numerical models of GLAO
based on measurements of the vertical distribution of turbulence at a number of sites
\citep{RigautGL,TokGLa,TokGLb}, suggest the potential for dramatic image
improvement in the near infrared. \citet[in preparation]{GeminiGLAO} have modeled GLAO
performance in detail, using vertical distributions
of turbulence seen at Cerro Pach\`on that put roughly 2/3 of the power in the
ground layer. With a pentagonal arrangement of LGS and an adaptive secondary mirror,
the study concluded that the median image width in K band would improve from
$0\farcs42$ to $0\farcs18$ for a $5\arcmin$ field and to $0\farcs25$ over $10\arcmin$.  

Ground-breaking work by \citet{RRnature} has hinted at the power of tomographic
wavefront sensing.  That paper describes the first experiments in tomography,
at the Telescopio Nazionale
Galileo, in which wavefront estimates were obtained simultaneously from a close
asterism of four natural stars through the simple expedient of defocusing the
telescope.  The instantaneous Zernike modal amplitudes measured from one star were
compared to estimates derived from the modal amplitudes of the other three.
Tomographic estimation from the three beacon stars was shown to be substantially
superior to a simple average, or the use of any one of the three individually.

We report in this paper results from the first implementation of a wavefront
sensing system based on a constellation of multiple laser beacons.  Previous 
work on laser-guided AO has been confined to a single beacon, usually created by
resonance scattering in the sodium layer at 90 km altitude
\citep{LickLGS,SORLGS,KeckLGS}.  For our work at the 6.5 m MMT we have 
created five beacons, generated by Rayleigh scattering.  Commercial, pulsed
doubled-YAG lasers at 532 nm are used, readily available at moderate cost.
Single Rayleigh
beacons gated to an altitude of $\sim$ 12 km have previously been used for AO with
small telescopes \citep{SORRayleigh} but the strong focus anisoplanatism and
incomplete sampling of higher turbulence makes a single such beacon of little
value for an 8 m class telescope.  

For our experiments, though, Rayleigh beacons are ideal.  In
our system the range
gate is centered at 24 km altitude, high enough to sample collectively most
of the turbulence.
To increase signal strength, the sensor system includes optics to focus the telescope
dynamically, to maintain sharp images of each laser pulse at it rises from 20 to 29 km
\citep{61inch,Stalcup}.
In this way, the range gate is extended to be $\sim$ 20$\times$ longer than the
telescope's normal depth of field.  Even at this high altitude,
focus anisoplanatism would remain a difficulty with a single beacon, but with five,
it is not an issue, precisely because our goal is the development of
tomographic techniques.
The practical advantage is that the current high capital and
operating costs of tuned sodium lasers for resonance scattering are avoided,
allowing us to make an early start.

\section{Instrument description}
The multi-beacon experiment at the MMT has been designed to investigate both
full tomographic wavefront sensing needed for diffraction limited imaging and
ground layer sensing to improve the seeing. With these goals in mind, and
guided by previous modeling of tomography \citep{MLH02}, we have built a
pentagonal constellation of Rayleigh laser guide stars (RLGS)
with a diameter of $2\arcmin$.  This field is
not as large as one would choose for ground-layer sensing alone, given
theoretical predictions of the size of the field that might be corrected by a
GLAO system; nevertheless, it is capable of taking the first steps to
quantifying the degree of correction to be expected of a closed-loop system.

The experimental set-up divides in two parts: the lasers and beam projector
optics to generate the five RLGS, and the wavefront sensor detectors and
associated optics mounted at the telescope's Cassegrain focus.

\subsection{Generating the laser guide stars}
The mean range chosen for the Rayleigh return is 24 km, where the beacons
form a pentagon
14 m in diameter.  The collective volume filling of this arrangement allows
for three-dimensional sensing of the turbulence by tomography for ranges up
to 10 km, and thus avoid the focal anisoplanatism that would arise
from a single RLGS.

The projection system employs two frequency-doubled YAG lasers from
Lightwave Electronics, rated at a nominal 15 W output each, operating at
532 nm and 5 kHz pulse repetition rate
\citep{Stalcup}.  The two linearly polarized beams are combined with a
polarizing beam splitter to produce a single, diffraction-limited beam
measured at $\sim$ 27 W. In operation, synchronized pulses of light from the
lasers are split
by a computer-generated hologram in the projection optics to create the
5 beacons on a $2\arcmin$ diameter ring on the sky.  The lasers and combining
optics are attached directly to the 6.5 m telescope tube, with the power
supplies and chiller in an adjacent room in the co-rotating MMT building.
The laser heads are in a thermally controlled enclosure.
At the top of the telescope tube on one side
is a steering mirror close to an image of the exit pupil, used to compensate
for slow misalignments in the projected beam direction caused by flexure and
temperature changes. The hologram is also at this pupil. The beams are
launched together through a single 50 cm diameter telephoto objective lens
mounted centrally on the 6.5 m telescope axis, directly above the secondary
mirror. 

\subsection{Cassegrain instrument for wavefront sensing}

The wavefront sensor package mounts at the MMT's Cassegrain focus, and
includes distinct optical arms for the RLGS and a natural star.  The first
comprises a novel sensor, which includes the dynamic focus optics, described
in detail in \citet{61inch}, and a single electronically shuttered CCD
which images all five beacons.

We aim to measure laser returns from $\sim$ 24 km altitude, twice the height
of earlier Rayleigh beacons \citep{SORRayleigh}, to allow better sampling of
high turbulent layers.  Because the air at that altitude is only 10\% of the
sea-level density, we have developed a new technology  to focus the telescope
continuously to follow the rising pulse from a Q-switched laser
\citep{Angel,DRpaperMLH}.  The sensitivity is thereby increased by an order
of magnitude compared to conventional range gating methods. The dynamic focus
of all five beacons is achieved by sinusoidal motion of a single 25 mm mirror
placed at an image of the telescope pupil, with approximately one third of
each cycle matched to the motion of the returning laser pulse. 

The dynamic focus optics and mechanical resonator that moves the mirror are
included in the fore-optics of the RLGS sensor.  The resonator serves also as
the system's master clock, triggering the laser pulses and the RLGS wavefront
sensor (WFS) detector
shutter at prescribed phase delays.  In this way, each laser pulse is tracked
in focus over a range gate from 20 to 29 km.

The refocused laser light is imaged onto a common 35 mm pupil, where a prism
array \citep{Putnam} divides the light into 36 subapertures of equal size in
a hexapolar geometry.  A compound lens following the prism array images the
beacons into five separate 36-spot Shack-Hartmann patterns on the detector,
a CCID18 from Lincoln Laboratory.  This
is a $128\times128$ pixel frame transfer device with 16 output amplifiers and
an electronic shutter. The controller, supplied by SciMeasure, reads the
detector at frame rates up to 915 Hz.  Operation of the electronic shutter
during transitions of the CCD clock signals introduces additional noise into
the image, so a carefully interleaved sequence is required of the clocks
for the frame transfer and read operations, and of the shutter, which
is triggered by the dynamic focus oscillator.

In order to explore the effectiveness of the RLGS wavefront sensor as a function
of position within the field, a second optical arm of the Cassegrain instrument
was built with a standard Shack-Hartmann sensor to measure wavefronts from
natural stars. A dichroic splitter covering a $3\arcmin$ field separates the
beams to the two wavefront sensors, transmitting the laser light to the LGS arm
and reflecting starlight at wavelengths longward of 650 nm to the star sensor.

The star WFS is a copy of the one normally used for adaptive optics with the
MMT adaptive secondary mirror \citep{Guido}. The pupil is divided into a
$12 \times 12$ array of square subapertures of which 108 are illuminated.
The camera uses an E2V CCD39 frame transfer detector with $80 \times 80$
pixels.  The whole
WFS and associated optics were placed on a mechanical slide, allowing
translation in one axis across the $2\arcmin$ field covered by the beacons.
In practice, the telescope is offset a prescribed angle from the nominal
stellar coordinates, and the WFS is moved by a corresponding angle in order to
reacquire the star.

In operation, both sensors are run simultaneously, but asynchronously, with
the latter providing measurements of the true stellar wavefront aberration to
which reconstructions derived from the recorded RLGS signals can be compared.
The star WFS was typically run at its maximum unbinned frame rate of 108
frames per second, with the RLGS sensor running at approximately half that
frame rate.  Data capture was externally registered in time by the use of
synchronized flashing LEDs placed in front and to the side of each camera,
running at approximately 1 Hz.

\section{Observations and analysis}
\subsection{Photometric and imaging performance of the RLGS wavefront sensor}
We have quantified the return flux from the RLGS using observations of
the spectrophotometric standard star HD192281 made through a filter of
3 nm equivalent width, centered on 532 nm. To do so, the prism array defining
the subapertures in the RLGS WFS was replaced with a flat optic of the same
construction, and the telescope refocused to account for the substantial
difference in stellar and beacon conjugates. The detected quantum efficiency
of the WFS, including the telescope, optics and detector, was found to be 15\%.
Throughput of the beam projector optics was estimated at 73\% from measurements
of each element individually.

During observations of the RLGS, the detected
return flux at zenith was $1.1\times 10^5$ photoelectrons m$^{-2}$J$^{-1}$ over
a range gate of
20--29 km.  Accounting for our estimated efficiencies in both the outgoing and
incoming beams, this amounts to $1.0\times 10^6$ ph m$^{-2}$ at the telescope
aperture per joule at the output of the beam projector.  The expected
flux has been
calculated from the lidar equation \citep[p. 222]{hardy}, and the 
atmospheric transmission \citep[p. 264]{AQ}, taking into
account the atmospheric temperature and pressure profile measured by a
routine meteorological balloon flight from nearby Tucson International Airport
at approximately the same time as the observations.  The agreement, within
2\%, is good; fortuitously so, since the uncertainty in our
estimate of the projected power is 10--20\%, dominated by uncertainty in
the calibration of the lasers' built-in power meters.  The return flux
is as bright as a star of $m_v = 9.9$ seen through a V filter,
and is similar to the range of returns from sodium beacons: \citet{jian},
for instance, measured $1.2\times 10^6$ ph m$^{-2}$J$^{-1}$.

An unusual problem with the CCD in the RLGS WFS prevented the formation of
sharp images. The typical full width at half maximum (FWHM) of the
Shack-Hartmann spots was found to be 4.4 pixels (equivalent to $3\farcs7$)
when images of the beacons recorded independently on a separate acquisition
camera were measured to have a FWHM of $\sim 1\farcs5$. 
A replacement chip has since been installed that shows no sign of this
blurring.  The spot separation,
set by the WFS optics, was just $4 \arcsec$, so that there was substantial
overlap of neighbouring spot images.  An iterative deconvolution algorithm
was required to recover useable spot positions.  Furthermore, the increase
in width reduced the signal-to-noise (S/N) ratio of the determination of each
spot's position by a factor of $\sim 2.5$.  An examination of the power
spectral densities of the reconstructed modal amplitudes from the RLGS
showed that frequencies above 15 Hz were dominated by noise rather than
atmospheric signal.  In the following analysis, therefore, a low-pass filter
with a sharp cut at 15 Hz was applied to the data.

\subsection{Comparison of laser and natural star wavefronts}
\begin{figure}
\epsscale{0.8}
\plotone{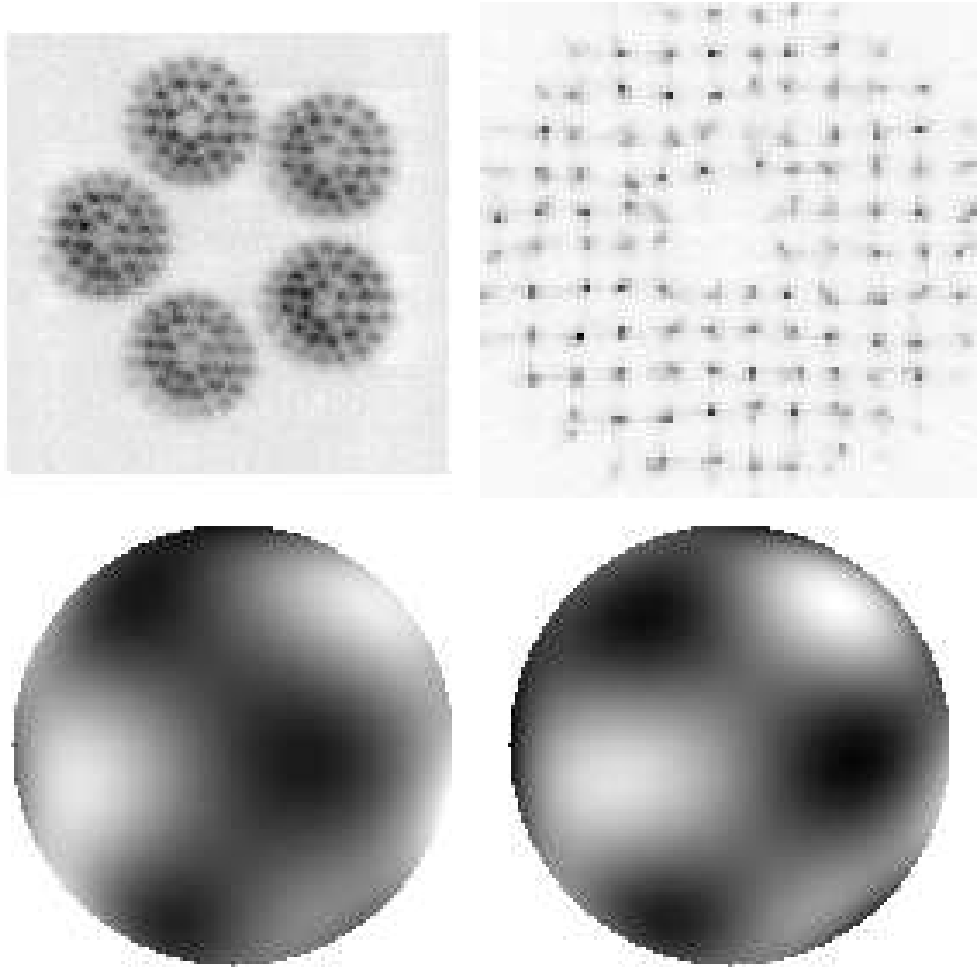}
\caption{Examples of the WFS data. The top panels show the WFS outputs,
with the RLGS sensor on the left and the star WFS on the right.  The
bottom panels show the corresponding wavefronts reconstructed to Zernike
order 6. The RLGS reconstruction is the average of the five individual
wavefronts.}
\label{wfsfig} 
\end{figure}

Wavefront data were recorded during a 2 hr period on 2004 September 30,
during which the seeing, $1\farcs0$ in V band, was worse than median for the
telescope.  The recorded data are frames from the
RLGS and star WFS cameras, in near-continuous sequences of 9 s. Figure
\ref{wfsfig} shows sample frames taken simultaneously from the two
cameras, and the corresponding reconstructed wavefronts, which in the
case of the RLGS is the average of the five individual reconstructions.
Both reconstructions are complete for the 25 modes in Zernike radial
orders 2 through 6, although the resolution of the star wavefront
sensor would allow more modal amplitudes to be calculated.  Overall
tip and tilt are of course not recovered from the lasers because of
unknown beam motion in the upward paths.

\begin{figure}
\epsscale{0.5}
\plotone{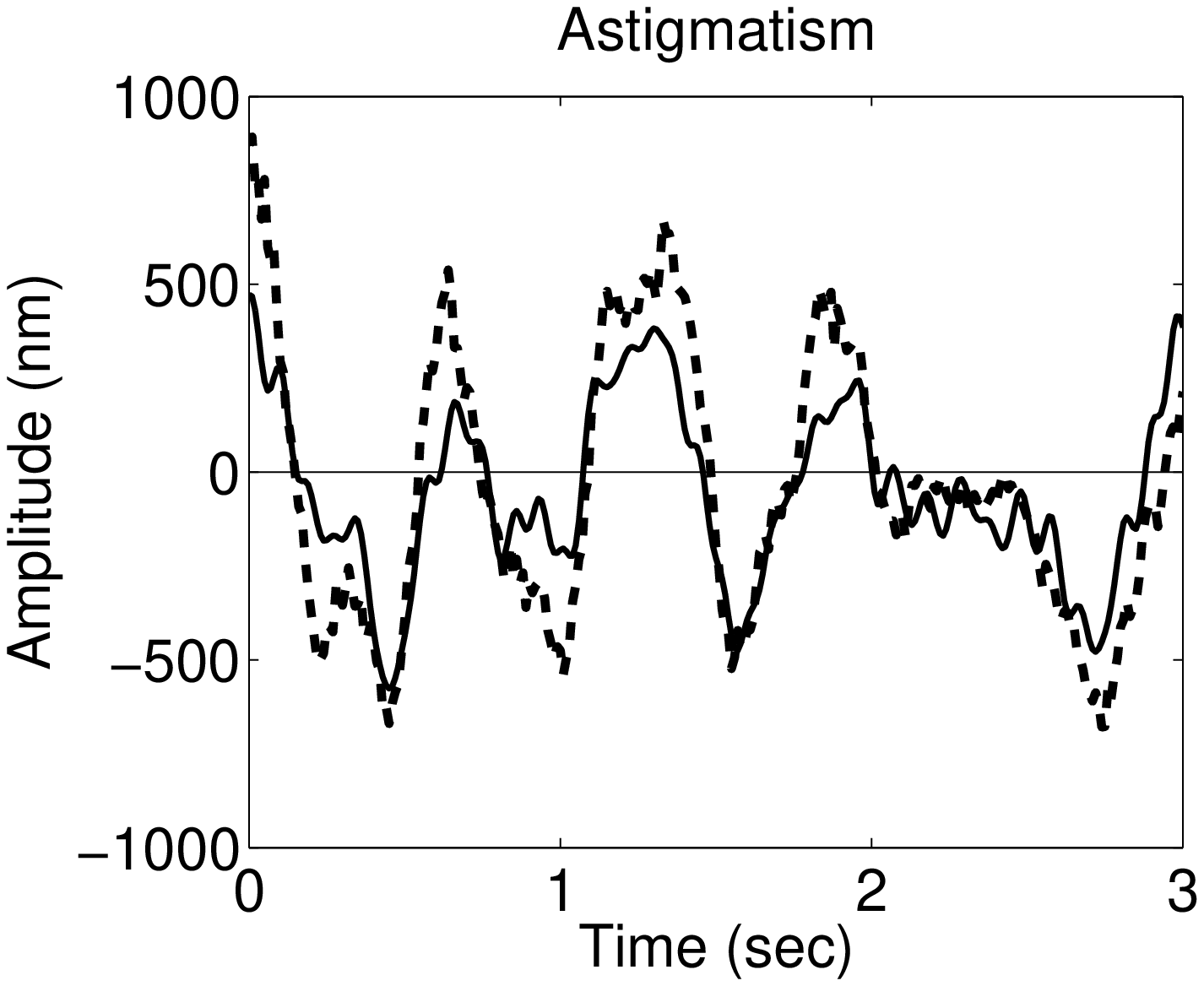}
\plotone{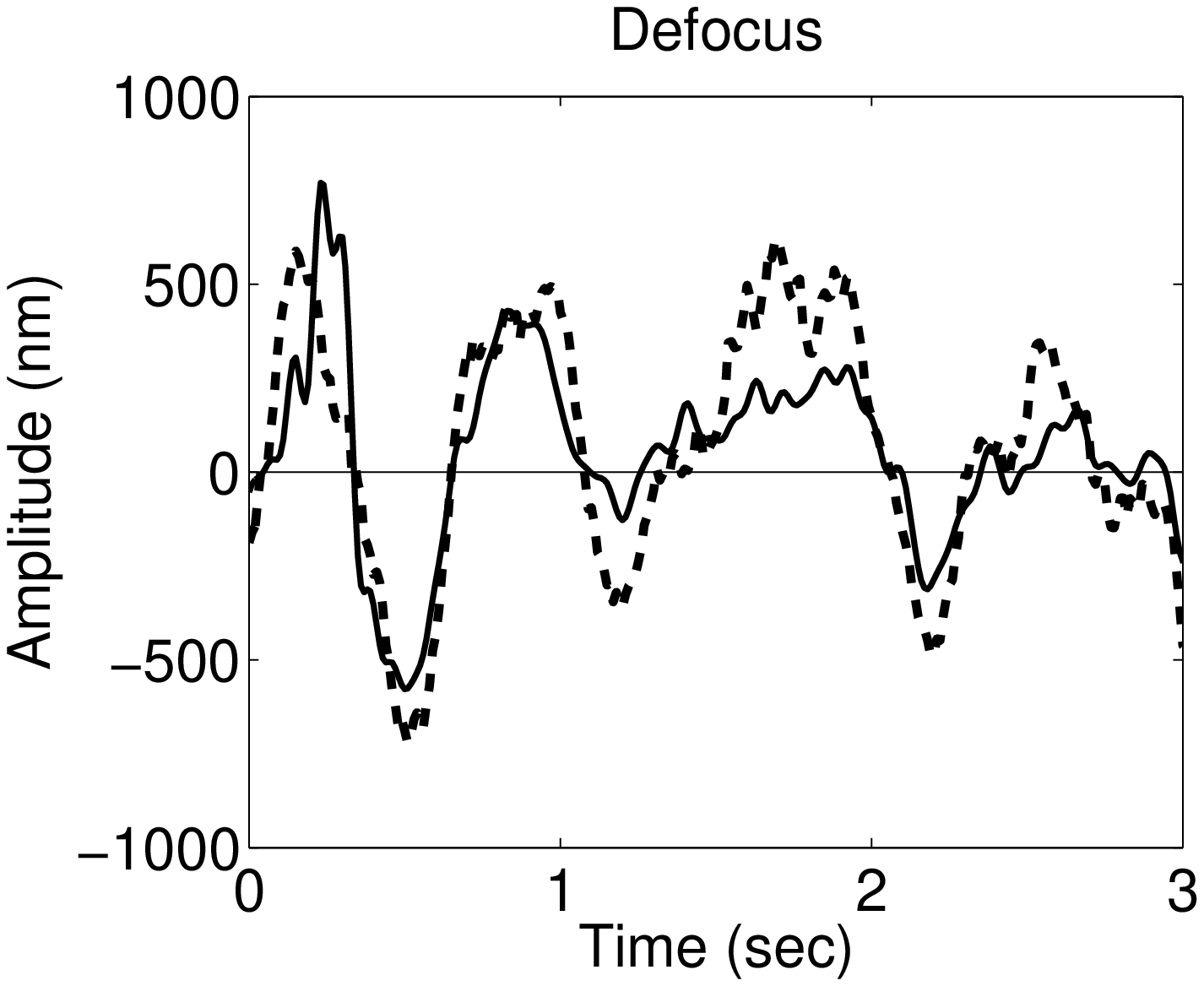} 
\plotone{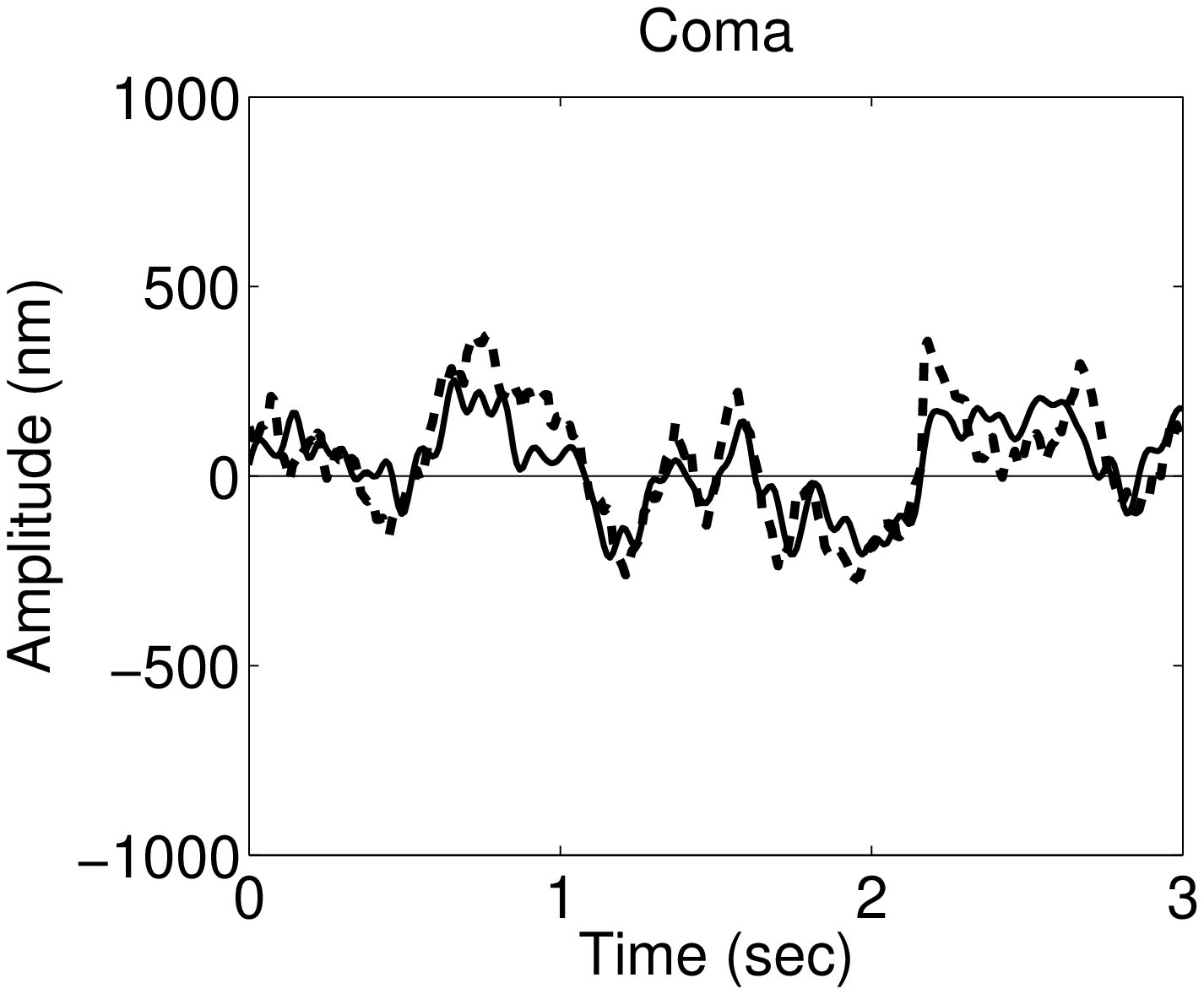}
\caption{Examples of the reconstruction of three Zernike modal amplitudes from
the natural star (dashed line) and the average RLGS signal (solid line).}
\label{zamps}
\end{figure}

Reconstructions of three individual modes as they evolve in time over a
period of 3 s are shown, by way of example, in Figure \ref{zamps}.  Values
for the modal amplitudes for the average of the five beacons and
the simultaneous values from a star near the center of the RLGS field are
plotted.  The generally good agreement indicates the presence of a strong
ground layer.  Aberration very close to the telescope's aperture will be common
to all five RLGS and the star, and so will contribute in the same way to the
variation of the modes.  On the other hand, high-altitude turbulence will be
sampled differently by the six light beams, and leads to the differences in
the recovered amplitudes.

\begin{figure}
\epsscale{1.0}
\plotone{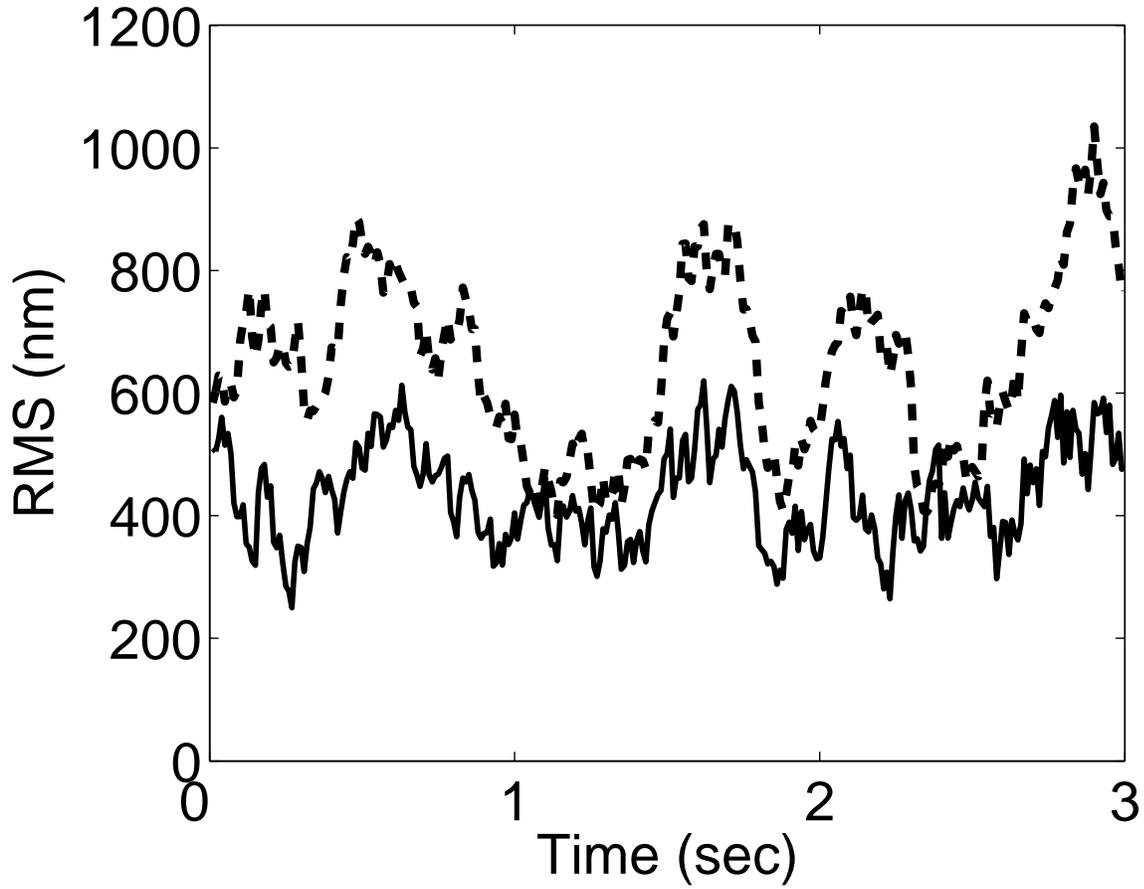}
\caption{Wavefront aberration averaged over the 6.5 m pupil for a
representative sample of data.  The dashed line shows the total for
the reconstructed modes in Zernike orders 2 through 6 in the stellar
wavefront.  The solid line shows the residual after subtraction of
the average RLGS wavefront.}
\label{temporal}
\end{figure}

\begin{figure}
\epsscale{1.0}
\plotone{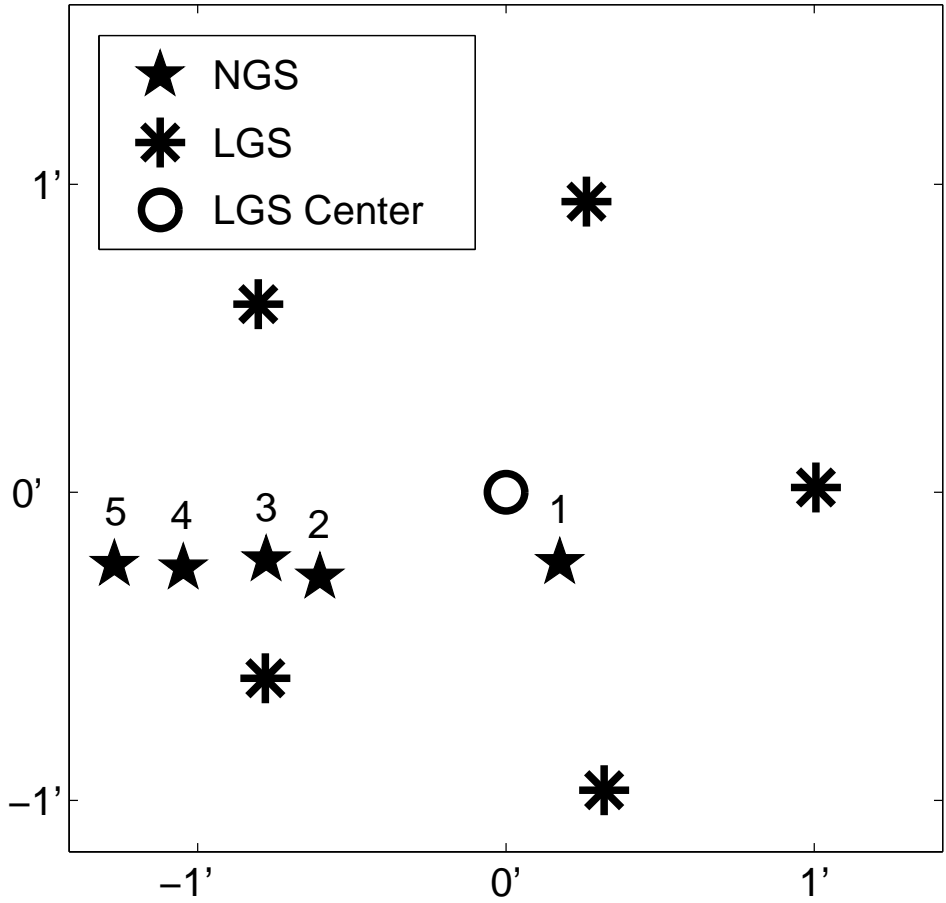}
\caption{Geometry on the sky of the RLGS used to measure the effect of
boundary layer wavefront sensing.  The labeled positions of the natural
star where data were recorded are separated from the geometric center
of the RLGS constellation by respectively $17\farcs2$, $39\farcs8$,
$48\farcs5$, $64\farcs5$, and $77\farcs5$.}
\label{geometry}
\end{figure}

By subtracting the average RLGS wavefront from the stellar wavefront, one
can obtain a quantitative  estimate of the degree of correction possible with a simple
ground-layer correcting scheme in which a DM conjugate to the
telescope pupil is driven in response to the averaged beacon signals.
Figure \ref{temporal} shows the result for a 3 s sample of data.  In general,
we find that both the mean and the variance of the aberration are substantially
reduced.  We have investigated the potential for such correction for stars
at several locations across the RLGS field.  The geometry of  the beacons and
the positions of the stars in the field used for this study is laid out in
Figure \ref{geometry}.

For the five positions of the star, Table 1 shows the measured stellar
wavefront aberration averaged over all the Zernike modes in each order, both
before and after subtraction of the average RLGS wavefront.  The value of
$r_0$, deduced from the image motion of the Shack-Hartmann spots in the
stellar WFS, and scaled to 500 nm wavelength, is also shown.  The overall
degree of correction varies from 40\% near the center to 33\% just outside
the field covered by the RLGS.

\begin{figure}
\plotone{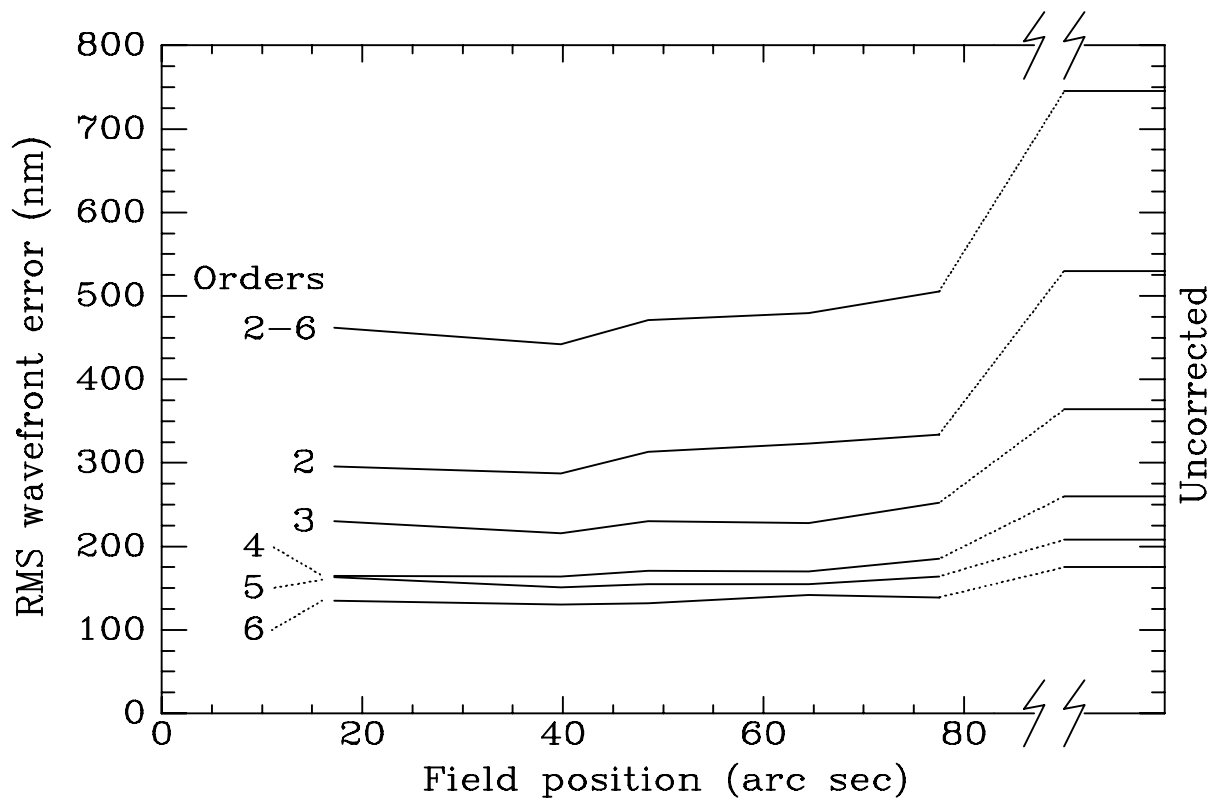}
\caption{Graph of the rms residual error in the stellar wavefront after subtraction
of the average beacon wavefront measurement.  The plotted lines show the
residual in all the modes of Zernike radial orders 2 through 6,
averaged over time, and the contributions of each order separately.
The short lines on the right represent the uncorrected rms wavefront
aberrations.  The points at each field angle
represent averages of 9 s of data, and in each
case the data have been scaled to the mean $r_0$ seen during the observations
of 10.1 cm at 500 nm wavelength.}
\label{wferror}
\end{figure}

Because the seeing varied significantly over the 2 hr period spanned by
these observations, a direct comparison of wavefront correlation versus
field angle is possible only by scaling the results to a common value.
Figure \ref{wferror} shows the results graphically, where the rms wavefront
values in Table 1 have been scaled by a factor $(r_0/10.1~{\rm cm})^{5/6}$, to
the average value of $r_0$ of 10.1 cm seen during the observations.  The
greatest improvement is seen in the lowest orders, which is to be expected,
both because the amplitudes of the lower order modes have larger variance and
are therefore sensed with greater S/N ratio, and because their angular
correlation scale is larger and therefore correction extends to higher
altitude than for modes with high spatial frequency.  We note also that
random noise in the reconstructions will preferentially appear as
differences amongst the five RLGS signals, and so our estimate of the
correlation between the  natural star and mean LGS wavefronts is conservative.

\subsection{Vertical distribution of turbulence}
In a simplified model in which the vertical distribution of turbulence is
characterized by just two regimes, one near the ground that is well corrected,
and the other at high altitude that is not corrected, one can readily calculate
from our data the division of power in the aberration between the two. Under the
assumption of a Kolmogorov spectrum, with an infinite outer scale, the mean
square wavefront error in orders 2--6 is $0.118 (D/r_0)^{5/3}$ rad$^2$
\citep[p. 96]{hardy}. A deficit of power is observed in the two tilt modes
seen by the NGS WFS, which we attribute to a finite outer scale, but
in second and higher orders, the effect is small. In these orders, the overall
average ground-layer corrected residual error inside $70\arcsec$ radius is
464 nm, after the scaling to the mean seeing condition, yielding a value of
$r_0^{FA}$ for the uncorrected free atmosphere of 21.7 cm.  The ground layer,
then, is characterized by $r_0^{GL} = 12.3$ cm.  This division of power of
approximately 2/3 in the boundary layer and 1/3 in the free atmosphere is
consistent with typical conditions at other sites \citep{GLstudies}.

The thickness of the boundary layer is an important parameter which sets
the corrected angular field of GLAO, and about which very little is known
from current site surveys.  Only recently have techniques been developed,
such as SODAR \citep{Skidmore} and SLODAR \citep{Wilson}, that address the
detailed structure of the lowest levels of turbulence.

The turbulence-weighted
mean height of the ground layer, $h^{GL}$, on the night of the observations
can be estimated from the anisoplanatic behavior of the star and RLGS
wavefronts.  Figure \ref{aniso} shows the mean square difference, as a
function of angular separation, between the stellar wavefront and the
simultaneous wavefront seen by four of the RLGS separately.  One beacon,
which gave wavefronts noticeably noisier than the other four, has been
omitted. All 20 pairings of
the five star positions and RLGS are plotted, after the same scaling
to uniform $r_0$ as above.  Also shown is a fitted curve of the form
$y = a + b~\theta^{5/3}$, the expected behavior of the anisoplanatic error.
The fitted value of $b$ is $42.5\pm 3.4~{\rm nm}^2~{\rm arcsec}^{-5/3}$,
where the uncertainty bounds an increase of 1 standard deviation in
the $\chi^2$ fit metric.
This corresponds to an isoplanatic angle for the sensed boundary layer
only of $\theta_0^{GL} = 20\pm 1\arcsec$ at 500 nm wavelength.  The isoplanatic
angle is related to $h^{GL}$ through the standard formula $\theta_0 = 0.314
\cos\zeta~r_0 / h$ \citep[p. 103]{hardy}, where $\zeta$ is the zenith angle. Taking
the mean value for these observations of $\cos \zeta = 0.95$, we find $h^{GL}
= 380\pm 15$ m for the boundary layer turbulence.

\subsection{Projected GLAO gain with single and multiple RLGS}
\begin{figure}
\plotone{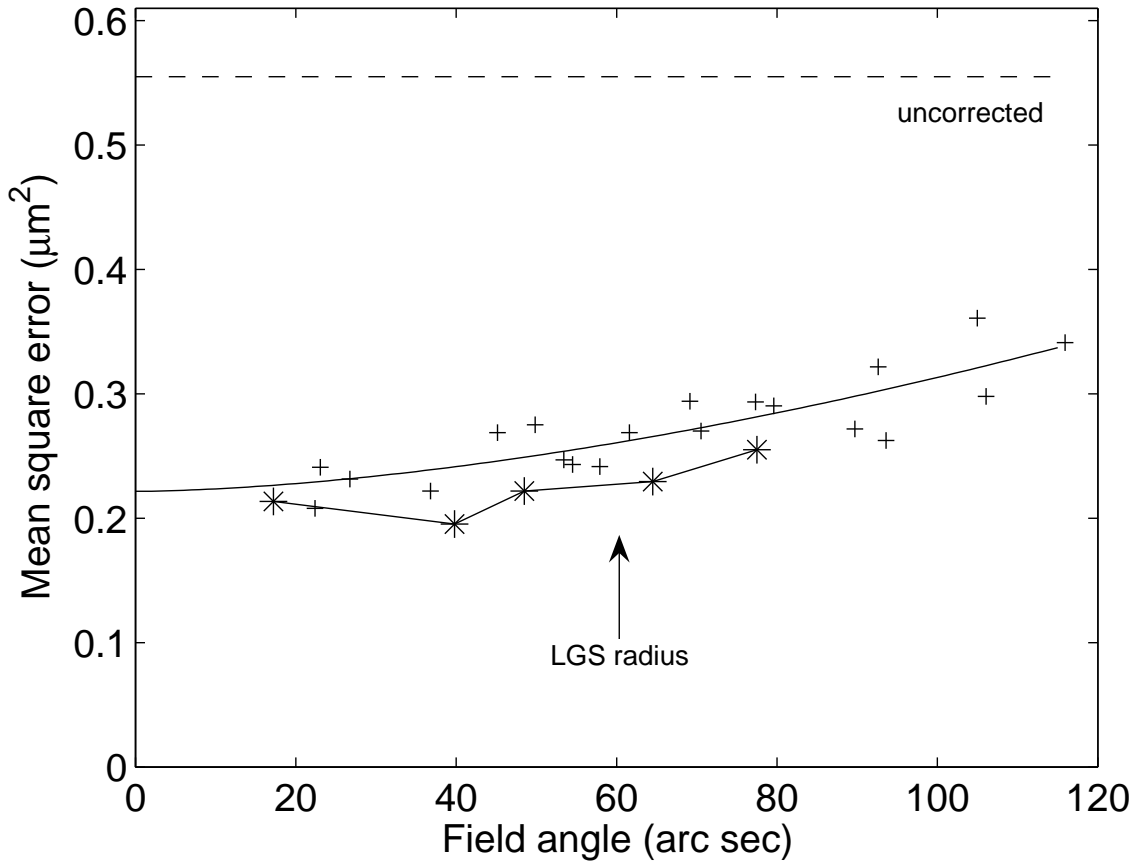}
\caption{The crosses show the mean square residual wavefront aberration after
correction of the stellar wavefronts by simultaneous wavefronts from
each of the RLGS, taken singly, as a function of the angular separation
of the star and beacon.  A fit to a 5/3 power law is shown as the solid
curve.  The stars reproduce the top line of figure \ref{wferror}, showing
the correction obtained from the average of the beacon wavefronts
from the same data.  The dashed line shows the mean square aberration in the
uncorrected stellar wavefronts.}
\label{aniso}
\end{figure}

Under conditions of turbulence dominated by the ground layer, represented
by our data, useful correction of seeing over a field of several arcminutes
would be obtained with a simple correction system using just one Rayleigh beacon.
By extrapolation of the curve of figure \ref{aniso} we find that the rms
aberration would be improved out to an angle, set by $\theta_0^{GL}$, of
$\sim~2\farcm5$ where the mean square residual caused by angular decorrelation
of the ground layer equaled the contribution from the uncorrected free atmosphere.
For such a simple correction system, the Rayleigh laser would be preferred over
a sodium laser, because it weights the ground layer more strongly than the
decorrelated higher
layers.  This is the approach adopted by the SOAR telescope \citep{SOARGLAO}.

Wavefront correction with multiple beacons will perform better over wide
fields.  Averaging the beacon signals leads to higher rejection of free
atmosphere aberration.  Correction of stellar wavefronts in the direction of
any given beacon will therefore be worse than if the signal from the beacon
were used on its own.  Over the field enclosed by the beacons, however, the
wavefront compensation is expected to be more uniform.  This effect, small
but appreciable for the narrow constellation diameter we used,
can be seen in the bottom trace in figure \ref{aniso}, where we superpose the
residual error after correction with orders 2--6 from the multi-beacon average,
from figure \ref{wferror}.  The improvement at $1\arcmin$ radius is 13\%.  

\section{Multi-beacon systems for current and future large telescopes}

\subsection{Practicality of multi-beacon systems}

Implementation of practical AO systems with even one laser beacon has proven
difficult. Sodium resonance systems were first proposed by \citet{Happer}, yet
are only now starting to be scientifically
productive. As a result, the concepts required for ELTs, which must rely on
still more complex systems with multiple beacons to realize their unique high
resolution potential, have been even slower to develop. 

Our first steps with a multi-beacon system give encouragement that the
engineering difficulties can be overcome. The laser wavefront sensor system
was designed to be as simple as possible, yet in its early stages, it has 
already proven capable of tracking the evolution of turbulence through
6th order Zernike modes over 6.5 m aperture.  After the development of full
tomographic reconstruction, correction to this degree in
a closed loop implementation of LTAO would yield the diffraction limit
at L band and longward, where the MMT is particularly powerful because of
the low thermal background afforded by its adaptive secondary mirror.

A closed-loop system with broad scientific application will require response
faster than the 15 Hz limit of the present data, and smaller subapertures
to provide correction at shorter wavelengths.  We plan to increase the resolution
of our RLGS WFS to 60 subapertures, for correction reaching down to H band.
Replacement of the CCD in that sensor to overcome the electronic
blurring is expected to allow a substantial, though as yet unquantified,
improvement. Photon noise does not set a practical limit; our measured flux
at the WFS of $1.1\times 10^5$ ph m$^{-2}$J$^{-1}$ scales to 470 photons
per subaperture per frame at the higher resolution, and a rate of 500 frames
per second, adequate for full correction in K band.  We envision also a further
factor of 2.5 increase in photon flux by use of one 15 W laser for each beacon,
which would allow operation through the H band. We note that the required
beacon fluxes do not scale with aperture.  Since the sodium beacons needed for ELTs
yield about the same photon flux per watt, they need be no more powerful.  

\subsection{Multiple beacon concept for the Giant Magellan Telescope}
The primary motivation for our work has been to understand the
implementation of multi-beacon AO for the 25 m GMT \citep{Johns}.
Like the MMT, the GMT will be equipped with an adaptive secondary mirror,
for large field GLAO correction and optimum performance in the thermal infrared.
Of particular value has been the realization that LTAO and GLAO can be considered
as the extremes of a continuum provided by beacon constellations of variable
diameter. The minimum diameter is that necessary for the beams to probe the
cylindrical volume traversed by light from a star on-axis, $1\farcm5$ for both
the MMT with Rayleigh beacons and the GMT with sodium beacons.  The maximum is
$\sim 10\arcmin$ for wide field GLAO.  In the multi-beacon implementation
at the GMT a single set of 5 beacons and 5 wavefront sensors will be configured
to cover this range of angles, in both the projection and sensing systems.  

The GMT's AO system will incorporate features proven valuable in the MMT prototype. 
For example, the constellation of beacons will rotate about the telescope axis in order
to counter field rotation.  This is done at the MMT by turning the hologram
that splits the single laser beam into five.
At the GMT, where we anticipate a separate laser for each SLGS, the same task
will instead be done with a K mirror.  The sensing system will be fed by reflection
from an $8\arcmin$ diameter dichroic mirror located above the direct Gregorian
focus, and will incorporate a set of 5 articulated periscopes to feed
light from the variable diameter constellation to fixed sensors.  Such
periscope optics are used now in the MMT, 
with a fixed reduction from $2\arcmin$ to $1\arcmin$.

Perspective elongation of the WFS spots will be significant, $\sim 2\farcs5$ in
the outer subapertures of the GMT.  Among several solutions proposed 
\citep{Beckers1,Beckers2,Ribak,Baranec}, our same dynamic refocus method at 5 kHz
proven at the MMT could be used also with pulsed sodium lasers.  The mechanical
requirements for dynamic refocus on the larger telescope are not substantially
more challenging than those already met at the MMT \citep{GMTAO}.

\subsection{RLGS constellations for 8 m class telescopes}

The question arises as to whether a Rayleigh beacon constellation might be used
as an alternative to a single or multiple sodium beacon for all AO operations at
an 8 m telescope.
While a single sodium beacon yields a good solution for the wavefront in the
direction of the beam, the corrected field is small, limited by the usual
angular anisoplanatism, and neither GLAO nor MCAO is possible.  Multiple
sodium beacons, the route being taken by the Gemini Observatory \citep{GeminiMCAO},
would add these capabilities, but at a high cost: the lasers are at present
more than an order of magnitude more expensive than YAG lasers of the same power.

Implementations of tomography with RLGS have been studied in simulation.
\citet{DRpaperMLH} modeled LTAO on an 8.4 m telescope with beacons range
gated from 16--30 km.  The corrected field of view remains small,
but the axial K-band Strehl ratio was 0.55, comparable to performance predictions
for 8 m telescopes with a single sodium beacon \citep{Viard}, and the values now
achieved in practice with a sodium beacon on the 3 m Shane telescope \citep{Lick}.
Although we have yet to simulate the comparison directly, we believe that it
may be possible for LTAO with Rayleigh beacons to improve on the performance
of a single sodium beacon.  This is because the beacons can be arranged to
provide 100\% sampling of the atmosphere up to a height above almost all
of the turbulence.  This contrasts with the single sodium beacon case, where
pupil coverage begins to drop immediately above the telescope, and falls off
with height.  A fraction of the turbulence is therefore unsensed, and more
importantly, an unknown radial shear is introduced into the measured
wavefront.

\begin{figure}
\plotone{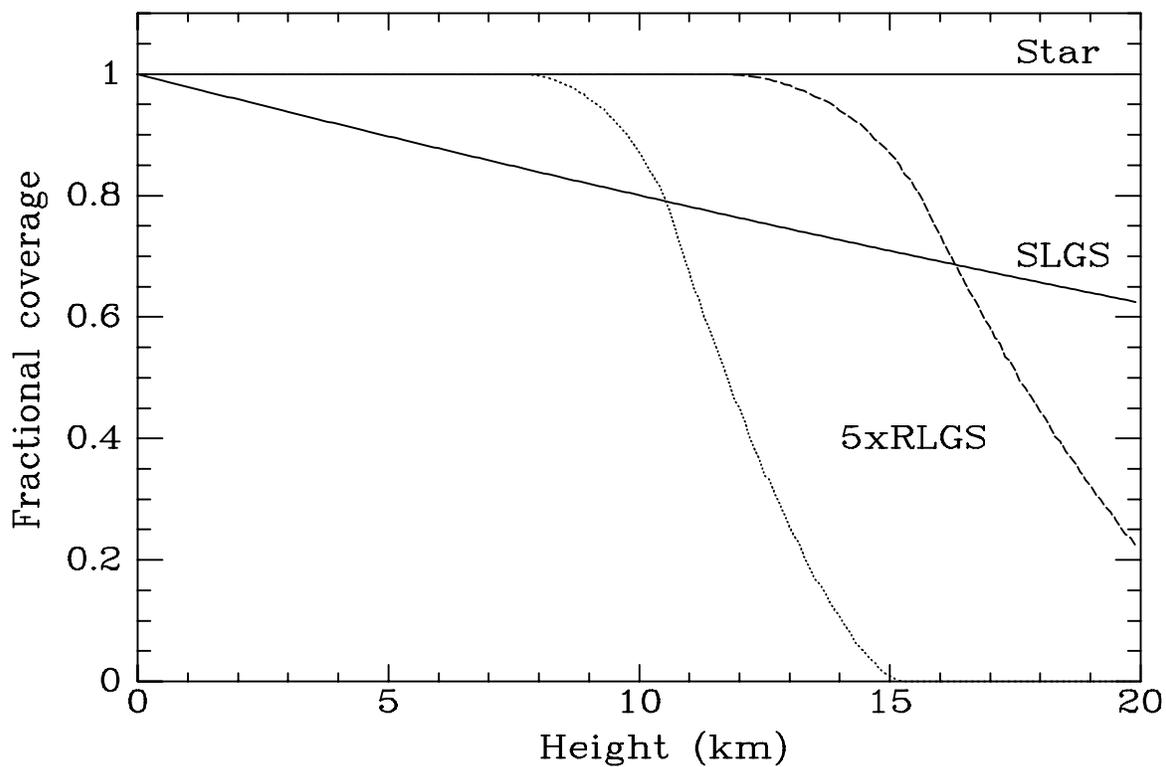}
\caption{Fractional intersection versus height of the area covered by light
from an axial star and three LGS beacon arrangements: a single axial sodium
laser guide star (SLGS) at 95 km range, and pentagonal constellations of Rayleigh
laser guide stars (RLGS).  We show the cases of a 6.5 m aperture, with the RLGS on
a $60\arcsec$ radius at 24 km range (dotted line), and an 8.0 m aperture with
RLGS at $43\arcsec$ radius and 30 km range (dashed line).}
\label{coverage}
\end{figure}

Figure \ref{coverage} illustrates the fractional coverage
of the area filled by on-axis starlight as a function of height by a single
sodium beacon at 95 km range, and a regular pentagon of Rayleigh beacons in
two cases.  For the present constellation diameter of $2\arcmin$ and mean height
of 24 km above a 6.5 m aperture, complete coverage is available to 8 km.
SCIDAR measurements above nearby Mt. Graham at the site of the Large
Binocular Telescope \citep{SCIDAR} indicate that this is above typically
75\% of the integrated $C_n^2$.  Also shown is the case of a telescope of
8.0 m diameter, where the beacon diameter has been set to $86\arcsec$, which
gives the greatest height for full coverage, and the beacons projected to a mean
range of 30 km.  There, full coverage extends to 12 km.  Even above that
height, the Rayleigh beacons continue to provide better atmospheric
sampling than the sodium up to 16.3 km, high enough to capture essentially
100\% of the integrated $C_n^2$.  Furthermore, tomographic reconstruction
from the multiple Rayleigh beacons can in principle remove much of the
focal anisoplanatism error inherent in the single sodium beacon measurement.

\citet{MLH02} have simulated the performance of MCAO on the
MMT using two DMs and five Rayleigh beacons.  They find that correction to the
diffraction limit in the near IR is achieved over a $1\arcmin$ field, if the
RLGS are each sampled at two distinct heights.  Modeling by \citet{delarue},
comparing MCAO systems on an 8 m telescope relying on Rayleigh beacons,
sodium beacons, and a combination of the two, came to the same conclusion.
That study shows further that the performance of the doubly-sampled Rayleigh
system approaches that of sodium beacons in the same geometry.
In practice, such a system of Rayleigh beacons
could be built with polarized lasers and a Pockels cell to switch
each returning beam from one WFS to another as each laser pulse rose through the
atmosphere.  The lasers, optics, and detectors to implement the scheme
would be still much cheaper than the lasers needed to construct an equivalent
with sodium lasers.

Rayleigh beacons at 30 km of similar brightness to those at 24 km at the MMT
could be made with individual lasers of 15 W.  The results of the models
described above suggest that a system on an 8 m telescope
that deployed five 15 W lasers, each dedicated to a single beacon,
would allow the full range of LTAO, MCAO, and GLAO techniques to be implemented.

\section{Conclusions}
Our results show that a constellation of RLGS will be a powerful
tool for ground-layer adaptive correction, when used to drive the MMT's adaptive
secondary mirror.  Other large telescopes equipped with adaptive secondaries
and with similar boundary layers of $\sim 500$ m mean thickness could also
benefit with much improved K-band images expected from GLAO over a field as
large as $10\arcmin$ diameter.

The main consideration that remains unexplored so far by our experiments is
the power of tomographic methods applied to multiple Rayleigh beacons to
recover higher layer aberrations as well as the ground layer.
Open-loop field tests to show the practicality of tomography are planned with the
MMT system.  Wavefronts will be reconstructed as Zernike modes on two
metapupils at the ground and at high altitude.  The choice of the upper
altitude is not critical \citep{MLH03}.  The performance
of LTAO will be evaluated by comparing direct measurements of natural star
wavefronts in the center of
the RLGS constellation with the integral of the tomographic solution
along the same line of sight.  In addition, the two-layer reconstructions
will allow an early investigation of the performance of MCAO by exploring
the agreement of the stellar and the integrated reconstructed wavefronts
over the field enclosed by the RLGS.

Our plans also call for closed-loop demonstrations of both
GLAO and LTAO with the MMT's adaptive secondary mirror,
and the addition of a new real-time reconstructor computer recently received
from Microgate S.r.l.  Later, the addition of a second DM conjugated to high
altitude will allow us to explore MCAO.

\acknowledgements
This work has been supported by the National Science Foundation
under grant AST-0138347
and the Air Force Office of Scientific Research under grant F49620-01-1-0383.
We are grateful for the
continued support of the MMT Observatory staff, particularly A. Milone,
J. McAfee, and M. Alegria.

\clearpage

\begin{deluxetable}{clllll}
\tablecolumns{6} 
\tablewidth{0pc}
\tablecaption{Wavefront Aberration Before and After Correction}
\tablehead{
\colhead{Zernike order}    &  \colhead{Set 1} & \colhead{Set 2} &
\colhead{Set 3} & \colhead{Set 4} & \colhead{Set 5} }
\startdata
2 & 462 & 572 & 513 & 571 & 559 \\
& {\it 255} & {\it 316} & {\it 308} & {\it 349} & {\it 343} \\
3 & 308 & 404 & 365 & 383 & 379 \\
& {\it 198} & {\it 238} & {\it 226} & {\it 246} & {\it 258} \\
4 & 223 & 285 & 261 & 276 & 269 \\
& {\it 142} & {\it 181} & {\it 168} & {\it 184} & {\it 190} \\
5 & 183 & 220 & 207 & 220 & 220 \\
& {\it 140} & {\it 166} & {\it 152} & {\it 168} & {\it 168} \\
6 & 159 & 184 & 175 & 194 & 170 \\
& {\it 116} & {\it 143} & {\it 130} & {\it 154} & {\it 143} \\
2--6 & 645 & 809 & 732 & 797 & 778 \\
& {\it 397} & {\it 487} & {\it 463} & {\it 518} & {\it 518} \\
\cline{1-6}\\
$r_0$ (cm) & 12.1 & 9.0 & 10.3 & 9.2 & 9.8 \\

\enddata
\tablecomments{The root mean square stellar wavefront aberration in nm,
averaged over the modes of each Zernike order, is shown before correction
by the average RLGS signal (Roman), and after correction (below, italic).
The sum over all measured orders is shown near the bottom, and the
last line reports the observed value of $r_0$ in cm at 500 nm
wavelength.}
\end{deluxetable}

\clearpage

\end{document}